\documentclass[%
 reprint,
 superscriptaddress,
 amsmath,amssymb,
 aps,
prb,
floatfix,
longbibliography 
]{revtex4-2}

\usepackage[pdftex]{graphicx} \graphicspath{{}}% Include figure files
\usepackage{float}
\usepackage{dcolumn}% Align table columns on decimal point
\usepackage{bm}% bold math
\usepackage[pdftex,colorlinks=true]{hyperref}
\hypersetup{
  colorlinks=true,
  linkcolor=blue,
  urlcolor=cyan,
}
\newcommand{\eqn}[1]{\begin{equation} #1 \end{equation}}
\newcommand{\eqa}[1]{\begin{align} #1 \end{align}}

\usepackage{color}
\definecolor{ForestGreen}{RGB}{34, 139, 34}

\newcommand{\amcr}[1]{\textsf{\color{red}(#1)}}

\newcommand{\nn}{\nonumber}

\newcommand{\mH}{\mathcal{H}}
\newcommand{\avg}[1]{\left\langle #1 \right\rangle}
\newcommand{\pd}{\partial}
\newcommand{\bS}{\boldsymbol{S}}
\newcommand{\bh}{\boldsymbol{h}}
\newcommand{\bx}{\boldsymbol{x}}
\newcommand{\bz}{\boldsymbol{z}}
\newcommand{\ep}{\mathcal{E}}
\newcommand{\mO}{\mathcal{O}}

\newcommand{\mC}{\mathcal{C}}
\newcommand{\mA}{\mathcal{A}}
\newcommand{\Qp}{Q^{\parallel}}
\newcommand{\Ep}{E^{\parallel}}

\DeclareMathOperator{\XY}{XY}
\DeclareMathOperator{\eq}{eq}
\DeclareMathOperator{\var}{var}
\DeclareMathOperator{\muth}{th}
\DeclareMathOperator{\csch}{csch}
\DeclareMathOperator{\eff}{eff}
\DeclareMathOperator{\KPZ}{KPZ}

\begin{document}

\preprint{APS/123-QED}

\title{Anomalous Dynamics and Equilibration in the Classical Heisenberg Chain}% Force line breaks with \\

\author{Adam J. McRoberts}
\affiliation{Max Planck Institute for the Physics of Complex Systems, N\"{o}thnitzer Str. 38, 01187 Dresden, Germany}

\author{Thomas Bilitewski}
 \affiliation{Max Planck Institute for the Physics of Complex Systems, N\"{o}thnitzer Str. 38, 01187 Dresden, Germany}
 \affiliation{JILA, NIST, and Department of Physics, University of Colorado, Boulder, CO 80309, USA}
 \affiliation{Center for Theory of Quantum Matter, University of Colorado, Boulder, CO 80309, USA}
 
\author{Masudul Haque}
 \affiliation{Max Planck Institute for the Physics of Complex Systems, N\"{o}thnitzer Str. 38, 01187 Dresden, Germany}
 \affiliation{Department of Theoretical Physics, Maynooth University, Co. Kildare, Ireland}
 \affiliation{Institut f\"ur Theoretische Physik, Technische Universit\"at Dresden, 01062 Dresden, Germany}

\author{Roderich Moessner}
 \affiliation{Max Planck Institute for the Physics of Complex Systems, N\"{o}thnitzer Str. 38, 01187 Dresden, Germany}

 %\email{amcr@pks.mpg.de}
 
\date{\today}% It is always \today, today,

\begin{abstract}
\noindent
% Abstract is rewritten based on new emphasis.
The search for departures from standard hydrodynamics in many-body systems has yielded a number of promising leads, especially in low dimension. Here we study one of the simplest classical interacting lattice models, the nearest-neighbour Heisenberg chain, with temperature as tuning parameter.  Our numerics expose strikingly different spin dynamics between the antiferromagnet, where it is largely diffusive, and the ferromagnet, where we observe strong evidence either of spin superdiffusion or an extremely slow crossover to diffusion. This difference also governs the equilibration after a quench, and, remarkably, is apparent even at very high temperatures.

\end{abstract}

\maketitle

%%%%%%%%%%
%\paragraph{Introduction}
%%%%%%%%%%
{\it Introduction.---}%
% should we add sections again, since it's no longer a PRL?
%\section{Introduction}
Hydrodynamics has long been a cornerstone of our understanding of many-body systems, and has recently become the focus of renewed inquiry. 
Hydrodynamic phenomena of interest in low-dimensional quantum systems include equilibration  \cite{lux2014hydrodynamic, leviatan2017quantum}, anomalous diffusion and transport  \cite{ljubotina2017spin,ljubotina2019KardarParisiZhang, weiner2020hightemperature, fava2020spin, dupont2020universal, bulchandani2020superdiffusive, schubert2021quantum, richter2021anomalous, dupont2021spatiotemporal, gopalakrishnan2019kinetic,Ilievski_2018,Ilievski_2021,DeNardis_2021}, hydrodynamics and superdiffusion in long-range interacting systems \cite{joshi2021observing,Schuckert_2020},  fracton and dipole-moment conserving hydrodynamics \cite{PhysRevResearch.2.033124,PhysRevLett.125.245303,PhysRevB.101.214205}, generalised hydrodynamics in integrable quantum systems \cite{Bernard2016,castroalvaredo2016emergent, bertini2016transport, bulchandani2017solvable,Sotiriadis_2017, DeNardis_2017,Doyon_2018,doyon2018soliton, bulchandani2018bethe, doyon2019lecture,Dubail_2020,Wrighton_2020,alba2021generalizedhydrodynamic}, and weak integrability breaking \cite{Friedman_Gopalakrishnan_Vasseur_PRB2020, lopez2021hydrodynamics}. In addition, recent experimental studies are probing (emergent) hydrodynamics in interacting quantum spin models \cite{Zu2021,joshi2021observing,Jepsen2020,wei2021quantum}. Hydrodynamics in classical many-body systems in low dimensions also poses many questions, perhaps most notably the appearance of anomalous diffusion and anomalous transport, often attributed to the Kardar-Parisi-Zhang (KPZ) universality class \cite{KPZ_1986,Spohn_2006,Sasamoto_2010,Amir_2010,Kriecherbauer2010,Quastel_2015,VanBeijeren2012exact, Lamacraft_2013,Dhar_Spohn_FPU_PRE2014, Spohn_JStatPhys2014,Spohn_2014a,Spohn_2015,Spohn_2016, Sasamoto_2018,Lepri_Politi_PRL2020,Dhar_2019}. 

The focus of this work is the classical Heisenberg spin chain, for which the nature of hydrodynamics has provoked extensive debate.  Based on the lack of integrability, it has been argued that ordinary diffusion holds for both spin and energy \cite{gerling1989comment, gerling1990time, bohm1993comment, srivastava1994spin, oganesyan2009energy, bagchi2013spin, glorioso2020hydrodynamics}. However, there have also been proposals of anomalous behaviour \cite{muller1988anomalous, de1992breakdown, de1993alcantara, de2020universality}, including an argument for logarithmically enhanced diffusion \cite{de2020universality}. Ref.~\cite{glorioso2020hydrodynamics}, in contrast, has argued from a theory of non-abelian hydrodynamics that each component of the spin follows a separate, ordinary diffusion equation.

In this paper, we present a systematic numerical study of the dynamical correlations and equilibration dynamics over a wide range of temperatures, $T < |J|$ to $T = \infty$.  We find ordinary diffusion of both spin and energy at $T = \infty$ and ordinary diffusion of energy at all (nonzero) temperatures in both the ferromagnetic (FM) and antiferromagnetic (AFM) chains \cite{supplemental}.
Most strikingly, we find a qualitative difference between ferromagnetic and antiferromagnetic models at finite temperatures. This manifests as a \textit{temperature-dependent} finite-time dynamical exponent in the spin correlations of the ferromagnetic chain, which departs from the diffusive exponent $\alpha = 1/2$; whereas the  antiferromagnetic chain displays behaviour compatible with spin diffusion at all temperatures studied. 
This deviation is apparent even at high temperatures, where the correlation length is still of the order of a single lattice spacing -- far from the low-temperature regime where the distinction between quadratic ferro- and linear antiferromagnetic spin-wave spectra may play a role. 
We have thus identified a, hitherto perhaps unappreciated, fundamental difference between the dynamics of the FM and AFM models. 

The observed behaviour of the ferromagnet could be interpreted as anomalous diffusion with a temperature-dependent exponent, or alternatively as a crossover at remarkably large timescales, rendering diffusion in practice unobservable experimentally for a wide range of temperatures.  Intriguingly, at low temperatures where we obtain the best fit to a single power-law, we observe the KPZ exponent almost perfectly across three decades in time.  In addition, the spacetime profiles of correlation functions closely follow the KPZ scaling form.
This establishes intermediate time KPZ scaling at low temperatures in the FM Heisenberg model, even if ultimately followed by a crossover to normal diffusion at very long times.

As a related phenomenon, we study equilibration dynamics after quenches from an $XY$ to a Heisenberg chain. 
%
%We establish that the observed equilibrium exponents also determine the equilibration tails of observables. 
%
Equilibration  is shown to proceed via a power-law approach to the equilibrium value, with an exponent determined by that observed in the corresponding unequal-time equilibrium correlation function, again displaying anomalous finite-time exponents in the case of the FM.

%%%%%%%%%%
%\paragraph{Numerical Methods \label{Numerical Methods}}
%%%%%%%%%%
{\it Model.---}%
%\section{Model}
We consider the periodic-boundary classical Heisenberg chain, with Hamiltonian 
\eqn{
\mH = -J \sum_{i = 1}^{L} \bS_i \cdot \bS_{i + 1},  \quad \bS_1 = \bS_{L+1}, 
\label{H_H}
}
for unit length classical spins $\bS_i \in S^2$.  Here $J = 1$ for the FM chain, and $J = -1$ for the AFM chain. 
The dynamics is given by the classical Landau-Lifshitz equation of motion,
\eqn{
\dot{\bS_i} = \{\bS_i, \mH\} = J \bS_i \times (\bS_{i-1} + \bS_{i+1}),
\label{LLeom}
}
which we solve numerically \cite{supplemental}. 

In equilibrium, we probe the spin-spin correlations
\eqn{
\mC^S(j, t) = \avg{\bS_j(t) \cdot \bS_0(0)},
}
and the energy correlations
\eqn{
\mC^E(j, t) = \avg{E_j(t) E_0(0)} - \ep^2,
}
where $E_j = -J \bS_j \cdot \bS_{j+1}$ is the bond energy, and $\ep = \avg{E}$ is the internal energy density. We use internal energy and temperature interchangeably, via $\ep(T) = T - \coth(1/T)$ \cite{fisher1964magnetism, supplemental}. Also, the (equal-time) spin correlation length is
\eqn{ 
\xi(\ep) = -1/\log(-\ep),
\label{correlation length}
}
which, as a function of $\ep$, is the same for the Heisenberg (\ref{H_H}) and $XY$ chains (\ref{H_XY}) \cite{supplemental}. 

Both of these correlation functions are symmetric under parity and time-reversal. To evaluate these correlations for a given $\ep$, we first construct an ensemble of 20,000 initial states drawn from the canonical ensemble of $\mH$ at the temperature $T(\ep)$ \cite{supplemental, loison2004canonical}. Each state is evolved in time, cf. (\ref{LLeom}), with snapshots stored at intervals of $\Delta t = 10J^{-1}$. The correlation function at a fixed time difference $t$ is calculated by averaging over 1000 consecutive snapshots for every initial state.

%%%%%%%%%%
%\paragraph{Hydrodynamics and Scaling Functions \label{Scaling Functions}}
%%%%%%%%%%
{\it Hydrodynamics and Scaling Functions.---}%
%\section{Hydrodynamics and Scaling Functions}
The hydrodynamic theory posits an asymptotic scaling form for the correlations of the conserved densities,
\eqn{
\mC(x, t) \sim t^{-\alpha} \mathcal{F}(t^{-\alpha} x),
\label{scalingform}
}
with a scaling exponent $\alpha$ and universal function $\mathcal{F}$. 

The exponent is, in principle, independent of the precise form of $\mathcal{F}$, and may be extracted by fitting the autocorrelation function, $\mA(t) = \mC(0, t)$, to a power-law,

\eqn{
\mA(t) \sim t^{-\alpha}.
\label{power_law_scaling}
}

Ordinary diffusion corresponds to an exponent of $\alpha = 1/2$, and a Gaussian scaling function,
\eqn{
\mC(x, t) = \frac{\chi}{(\pi D t)^{1/2}} \exp \left[ - \left(\frac{x}{(D t)^{1/2}}\right)^2 \right],
\label{diffusive_scaling}
}
where $\chi = \int dx \,\mC(x, t) = \sum_j \mC(j, t)$, and $D$ is the diffusion constant. This scaling function may be obtained directly by solving the ordinary diffusion equation.

The most well-known anomalous scaling is the KPZ universality class, with an exponent $\alpha = 2/3$. There is no analytic form for the scaling function, but it is tabulated in \cite{prahofer2004exact}. 

Even if the the asymptotic behaviour is diffusive, one might have finite-time corrections.  The lowest order correction to a diffusive autocorrelation function from non-abelian hydrodynamics \cite{glorioso2020hydrodynamics} is of the form
\eqn{
\mA(t) \sim (Dt)^{-1/2} + \Lambda t^{-1}.
\label{crossover_power_law}
}
From finite-time data, it may be difficult to distinguish this behaviour from anomalous exponents \cite{lux2014hydrodynamic}. 

% It is quite generic that, for some finite region, it is possible to find parameter values such that the sum of two power-laws (\ref{crossover_power_law}) is in good agreement with a single power-law with a variable exponent (\ref{power_law_scaling}) - making the two scenarios difficult to distinguish - though it is perhaps natural to regard (\ref{crossover_power_law}) as the more plausible, \textit{a priori}, of the two. 

%%%%%%%%%%
%\paragraph{Equilibrium Correlations \label{Equilibrium Correlations}}
%%%%%%%%%%
{\it Equilibrium Correlations.---}%
%\section{Equilibrium Correlations}
\begin{figure}
    \centering
    \includegraphics[width=\columnwidth]{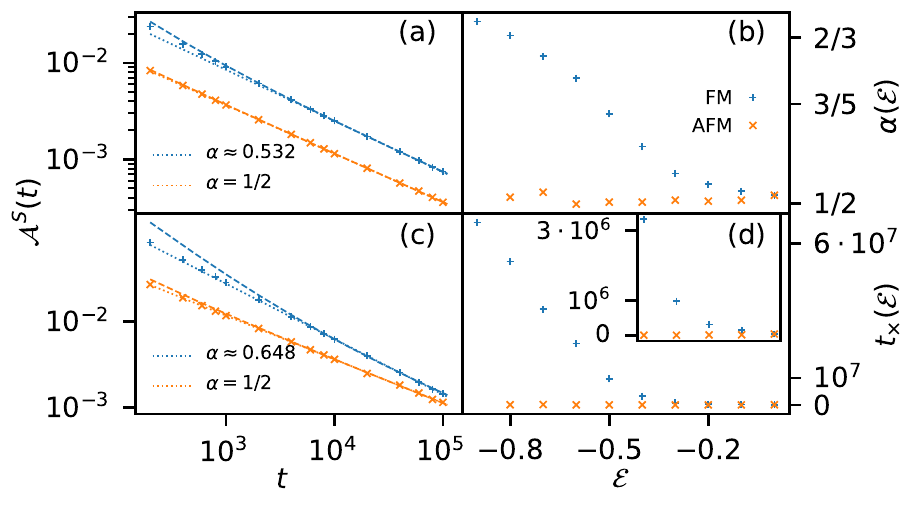}
    \caption{Anomalous hydrodynamics in the FM: power-law scaling of the autocorrelator $\mA^S(t)$, and extracted exponents and crossover scales for the FM (blue, $+$) and AFM (orange, $\times$). Panels (a) \& (c) show $\mA^S(t)$ for $\ep = -0.3$ \& $\ep = -0.7$, resp. The dotted lines show the power-law fit (\ref{power_law_scaling}), and the dashed lines show the finite-time corrected fit (\ref{crossover_power_law}).  Panel (b) shows the estimated anomalous exponents, while (d) shows the diffusion crossover times estimated from (\ref{crossover_power_law}) -- the inset zooms in on the points $\ep = -0.4$ to $\ep = 0$. 
    \label{fig:FM_vs_AFM}}
\end{figure}
We begin by examining the scaling exponent via the autocorrelation functions. We show $\mA^S(t)$ in Fig.~\ref{fig:FM_vs_AFM}(a) and \ref{fig:FM_vs_AFM}(c) for the FM and AFM at $\mathcal{E} = -0.3$ and $\mathcal{E} = -0.7$, for times $t = 200$ to $t = 10^5$.

The AFM displays ordinary spin diffusion, with the diffusive power-law observable after a comparatively short time $t \approx 10^3$. The FM does not exhibit diffusion, at any finite temperature, over the timescales of our simulations. 

The autocorrelations of the FM are, for these timescales, well-approximated by a power-law (\ref{power_law_scaling}), with superdiffusive exponents (Fig.~\ref{fig:FM_vs_AFM}(b)).  One may also fit a crossover of the form (\ref{crossover_power_law}).  Adopting this point of view, we may extract a crossover time $t_{\times}(\ep)$ after which we would predict the system to show diffusion, via the effective exponent
\eqn{
\alpha_{\mathrm{eff}}(t) = -\frac{d\log(\mA(t))}{d\log(t)},
}
with the crossover defined, arbitrarily, by $\alpha_{\eff}(t_{\times}) = 0.505$ (i.e., the time after which $\alpha_{\mathrm{eff}}$ is sufficiently close to $1/2$). The estimated crossover times obtained for the FM are orders of magnitude larger than the AFM, reaching $t_{\times} \approx 10^7 J^{-1}$ at low-to-intermediate temperatures (Fig.~\ref{fig:FM_vs_AFM}(d)).

\begin{figure}[t]
    \centering
    \includegraphics[width=\columnwidth]{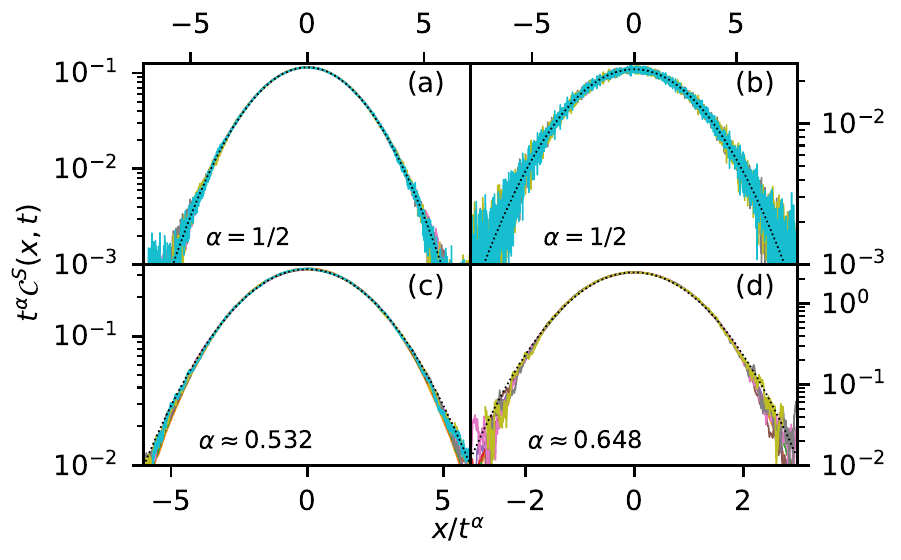}
    \caption{Scaling collapses of the spin correlations $\mC^S(x, t)$. Colours correspond to different fixed times from $t = 2000$ up to $t = 10^5$ ($t = 8\times 10^4$ in (d)). Panels (a) \& (b) show the diffusive collapse in the AFM at $\ep = -0.3$ and $\ep = -0.7$, resp.; (c) \& (d) show anomalous collapses at $\ep = -0.3$ and $\ep = -0.7$ in the FM, with exponents $\alpha = 0.532$ and $\alpha = 0.648$, resp. Dotted lines show a gaussian scaling function as a guide to the eye.
    \label{fig:scaling_collapses}}
\end{figure}

In Fig.~\ref{fig:scaling_collapses} we show the scaling collapses at these temperatures. The AFM is clearly consistent with a diffusive collapse (\ref{diffusive_scaling}); the FM is not. Since the autocorrelation function is well-fit by an anomalous power-law (\ref{power_law_scaling}), we use these exponents to perform the scaling collapse in the FM. This collapses the correlations rather well, though there is some noise in the tails. 

Moreover, despite the noise, one may observe that the tails of the correlations decay faster than a Gaussian. This suggests that an enhancement of the diffusion constant alone (whether of the form (\ref{power_law_scaling}), or a crossover (\ref{crossover_power_law})) is not the correct picture. 

\begin{figure}
    \centering
    \includegraphics[width=\columnwidth]{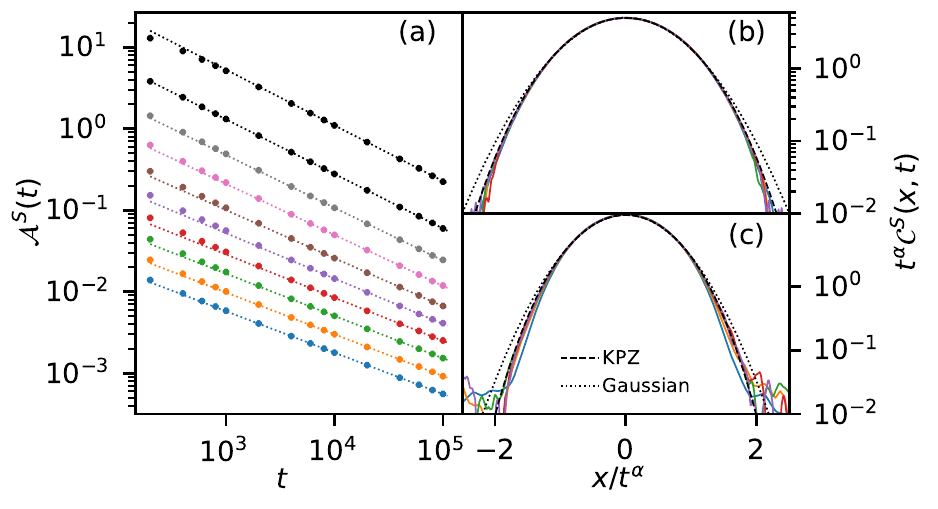}
    %\caption{ KPZ scaling in the FM. Panel (a) shows the inverse-width fit at $\alpha_{\KPZ} = 2/3$, and (b) shows the scaling collapse, at $\ep = -0.8$. Panels (c) and (d) show the same for $\ep = -0.9$.
    \caption{ Anomalous scaling in the FM. Panel (a) shows the the autocorrelator $\mA^S(t)$, vertically offset for clarity, for, in descending order, $\ep = -0.9$ to $\ep = 0$ in steps of $0.1$. The power-law fits have the exponents of Fig.~\ref{fig:FM_vs_AFM}(b), except the black lines at $\ep = -0.8$ and $\ep = -0.9$, which are fit with the KPZ exponent $\alpha_{\KPZ} = 2/3$. Panels (b) \& (c) show the scaling collapse using the extracted exponent comparing to the KPZ scaling function (dashed) and a gaussian (dotted) at $\ep = -0.8$ and $\ep = -0.9$, resp.
    \label{fig:KPZ}}
\end{figure}

%%%%%%%%%%
%\paragraph{KPZ Scaling \label{KPZ Scaling}}
%%%%%%%%%%
{\it KPZ Scaling.---}%
%\section{KPZ Scaling}
We thus examine the possibility that we are observing a crossover from KPZ scaling. Indeed, the numerical evidence at low temperature is remarkably strong, shown in Fig.~\ref{fig:KPZ}. The correlations up to $t = 10^4$ collapse onto the KPZ function for $\ep = -0.8$ and $\ep = -0.9$. Beyond this time the noise in the tails is too great to reliably distinguish the form of the spatial decay, but the scaling exponent measured by the autocorrelation function is consistent with $\alpha = 2/3$ up to the final time $t = 10^5$.
Moreover, at $\ep = -0.8$, there are apparently no finite-time corrections to the power-law decay $\mA(t) \sim t^{-2/3}$, for three decades in time. 

%In isolation, this would be strong evidence that the system is in the KPZ universality class - however, the fact that this is not the case at different temperatures traduces such a conclusion, and we expect that, towards infinite time, the system will eventually exhibit ordinary diffusion. 

%Irregardless of the ultimate infinite-time behaviour, the extremely long crossover -- and the existence of an intermediate time KPZ regime at all -- seems remarkable, and establishes a scenario beyond simple corrections to diffusive scaling in the FM Heisenberg chain.
%Despite this, we find it noteworthy that such clean evidence of the ``wrong" universality class persists numerically for such a long time.

%%%%%%%%%%
%\paragraph{Equilibration Dynamics \label{Equilibration Dynamics}}
%%%%%%%%%%
{\it Equilibration Dynamics.---}%
%\section{Equilibration Dynamics}
In addition to our equilibrium simulations examining unequal-time correlations, we perform equilibration simulations probing the relaxation to thermal equilibrium after a quench. This allows us to test whether the anomalous behaviour, and the distinction between FM and AFM, are also observable in out-of-equilibrium dynamics.

We initially prepare the system in a thermal state of the $XY$ chain, 
\eqa{
\mH_{\XY} = -J \sum_{i = 1}^{L} \bS_i \cdot \bS_{i + 1}
= -J \sum_{i = 1}^{L} \cos(\phi_i - \phi_{i + 1}),
\label{H_XY}
}
for unit length classical \textit{rotors} $\bS_i \in S^1$. At time $t = 0$, we quench the system, and evolve under the dynamics (\ref{LLeom}) of the Heisenberg chain. We examine the relaxation of the following observables: 
\eqn{
E^{\mu}(t) = -J\avg{S_i^{\mu}(t) S_{i + 1}^{\mu}(t)},
\label{E_mu}
}
which measures the energy attributed to the $\mu^{\muth}$ spin components; and
\eqn{
Q^{\mu}(t) = \avg{S_i^{\mu}(t)^2},
\label{Q_mu}
}
which measures the total magnitude of the $\mu^{\muth}$ spin components. These are natural measures of the anisotropy, which characterises the relaxation from the initial state, satisfying $S_i^z = 0\; \forall i$, to a (quasi-)thermal state of the isotropic Heisenberg chain. The equilibration of the energy fluctuations is measured using the heat capacity,
\eqn{
C(t) = \frac{\avg{\var{E_i(t)}}}{T^2},
}
where we take the spatial variance before the ensemble average to obtain a time-dependent quantity. 
As in the equilibrium simulations, we average over an ensemble of 20,000 states, initially drawn from the canonical ensemble of $\mH_{\XY}$. 

%In this case, the initial states are drawn from the canonical ensemble of $\mH_{\XY}$, at temperature $T_{\XY}(\ep)$ for a given $\ep$, where the correspondence is $\ep_{\XY}(T) = -I_1(1/T)/I_0(1/T)$, and $I_n$ denotes a modified Bessel function of the first kind. We fix $\ep$, not $T$, because only the former is unchanged by the quench.

We expect that the equilibration dynamics will be similarly hydrodynamic, since establishing the new global equilibrium requires the transport of conserved densities over long distances \cite{lux2014hydrodynamic}. The relaxation of an observable $\mO$ is therefore expected to follow a power-law
\eqn{
\delta \mO(t) := |\mO(t) - \mO_{\eq}| = \lambda t^{-\alpha},
\label{decay}
}
where $\mO_{\eq}$ is the thermal value of the observable in the Heisenberg chain. 

\begin{figure}
    \centering
    \includegraphics[width=\columnwidth]{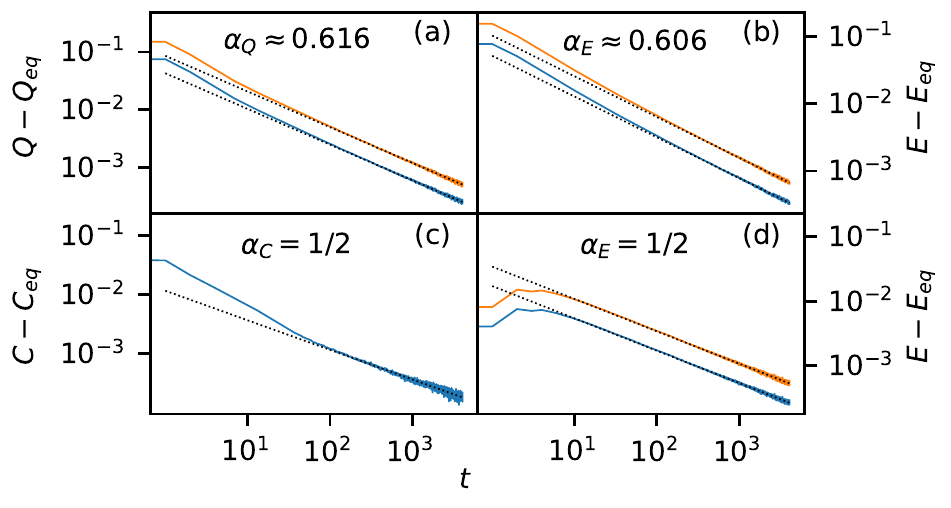}
    \caption{ Equilibration dynamics of $\mO^z(t) - \mO^z_{\eq}$ (blue) and $\mO^{||}(t) - \mO^{||}_{\eq}$ (orange), where $\mO^{||} = (\mO^x + \mO^y)/2$ is the average of the in-plane components. Panels (a), (b), and (c) show the equilibration of $Q$, $E$, and $C$ at $\ep = -0.5$ in the FM; (d) shows the equilibration of $E$ at $\ep = -0.5$ in the AFM. $Q$ and $E$ appear to equilibrate with an anomalous exponent of $\alpha \approx 0.6$ in the FM, though the data is equally well described by a combination of power-laws. The energy fluctuations (heat capacity) and the AFM equilibrate diffusively.
    \label{fig:equilibration}}
\end{figure}

These simulations exhibit complementary aspects of the same broad phenomenology observed in equilibrium. Fig.~\ref{fig:equilibration} shows the equilibration dynamics at $\ep = -0.5$.  The extracted (anomalous) equilibration exponents have qualitatively similar dependence on energy as those extracted from equilibrium correlation functions \cite{supplemental}.  The energy fluctuations, as measured by the heat capacity, always equilibrate diffusively. In the AFM, $E^{\mu}$ and $Q^{\mu}$ also show diffusive equilibration. In the FM, however, the equilibration of $E^{\mu}$ and $Q^{\mu}$ is anomalous. It should be noted that, although $E^{\mu}$ has dimensions of energy, it is, like $Q^{\mu}$, a measure of the magnetic anisotropy, and therefore equilibrates anomalously in the FM, rather than tracking the diffusive behaviour of the energy fluctuations.

Thus, as in equilibrium, we observe a striking difference between the FM and AFM, with only the former displaying anomalous exponents. 

While our simulations do not allow us to rule out a potential crossover to diffusive equilibration at even longer times, the observables can reasonably be described to have (fully) equilibrated with these anomalous exponents, in particular when considering a realistic experimental situation in which resolution and time scales might be limited.

%%%%%%%%%%
%\paragraph{Discussion \& Conclusions \label{Discussion and Conclusions}}
%%%%%%%%%%
{\it Discussion \& Conclusions.---}%
%\section{Discussion \& Conclusions}
We have conducted a detailed numerical study of the equilibrium and out-of-equilibrium dynamics of the classical Heisenberg chain, with the largest system sizes, simulation times, and range of temperatures we are aware of so far for this model.
We find that, although ordinary diffusion is expected at infinite time, the FM exhibits a long-lived regime that is well-described by an effective superdiffusive, temperature-dependent scaling exponent, with remarkably clean KPZ-like behaviour at low temperature. The AFM, by contrast, swiftly evinces ordinary diffusion at all temperatures.
The existence of such large intermediate scales is obviously relevant to experiments probing anomalous diffusion --- anomalous behaviour might be the only accessible experimental regime even when the longer-term behaviour is diffusive. 

A possible explanation of the intermediate-time regime and the stark difference between FM and AFM cases could be the presence of integrable ferromagnetic models, such as the continuum Landau-Lifshitz model \cite{Lakshmanan_PhysLettA1977_continuumHeisenberg,  Takhtajan_PhysLettA1977_continuousHeisenberg, Fogedby_JPhysA1980_continuumHeisenberg, book_FaddeevTakhtajan_1987_HamiltonianMethods} and the lattice FM model with $\log(1+\bS_{i}\cdot\bS_{i+1})$ interactions \cite{Ishimori_JPSJ1982, book_FaddeevTakhtajan_1987_HamiltonianMethods, Sklyanin_JournSovMath1988, Sklyanin_FuncAnal1982, Prosen_Zunovic_PRL2013, Dhar_2019}.  The Heisenberg FM studied in the present work could arguably be considered to be increasingly similar to either of these integrable models at lower temperatures.  This would be consistent with our finding that the low-temperature FM behaviour is closer to KPZ.  Recent work \cite{Ilievski_2018,gopalakrishnan2019kinetic,Ilievski_2021,DeNardis_2021} suggests the perturbative stability of KPZ scaling in systems which are close to integrability and preserve the rotational symmetry. 

Nevertheless, it is remarkable that we observe an anomalous regime even at near-infinite temperatures, where the correlation length (\ref{correlation length}) is short, e.g., already less than a \textit{single lattice spacing} at $\ep = -0.3$.  Intuitively, this regime does not seem to be close to either the continuum model or the integrable log-interaction model.  This points to the need for a better understanding of how proximity to integrable points might play a role in the physics of the Heisenberg FM, especially at high temperatures.

%\vspace{\baselineskip}

\def\CC{{\textsc{c}\nolinebreak[4]\hspace{-.05em}\raisebox{.4ex}{\tiny\bf ++}}}

\begin{acknowledgments}\noindent{\textit{Acknowledgements:}} 
A.J.McR wishes to thank J. N. Hallen, and P. Suchsland for helpful discussions, and particularly B. A. Placke for explaining various aspects of $\CC$. This work was in part supported by the Deutsche Forschungsgemeinschaft under grants SFB 1143 (project-id 247310070) and the cluster of excellence ct.qmat (EXC 2147, project-id 390858490). 
\end{acknowledgments}

\iffalse
{\it Appendix - Scaling Collapse Colour Map.---}%

\noindent
The map between colours and fixed times in all the scaling collapses shown in the main text is: t = 2000 (blue), t = 4000 (orange), t = 6000 (green), t = 8000 (red),\newline t = 10,000 (purple), t = 20,000 (brown),\newline t = 40,000 (pink), t = 60,000 (grey),\newline t = 80,000 (yellow), t = 100,000 (cyan).
\fi

% \nocite{mattis1984transfer}
\bibliography{refs} % Entries are in the "refs.bib" file

  %%%%%%%%%%% Merge with supplemental materials %%%%%%%%%%
  % Useful for the arxiv submission %
  % For the PRL submission the supplementary has to be submitted as a separate pdf anyway 
  \cleardoublepage
  \begin{center}
    \textbf{\large Supplementary Material}
  \end{center}
\setcounter{equation}{0}
\setcounter{figure}{0}
\setcounter{table}{0}
\makeatletter
\renewcommand{\theequation}{S\arabic{equation}}
\renewcommand{\thefigure}{S\arabic{figure}}
\renewcommand{\thetable}{S\arabic{table}}
\setcounter{section}{0}
\renewcommand{\thesection}{S-\Roman{section}}
%\renewcommand{\bibnumfmt}[1]{[S#1]}
%\renewcommand{\citenumfont}[1]{S#1}
%%%%%%%%%% Prefix a "S" to all equations, figures, tables and reset the counter %%%%%%%%%%

%%%%%%%%%%
\section*{Contents}
%%%%%%%%%%

In this supplementary we provide further details of our simulations, and further evidence for our conclusions. In \ref{Exact Thermodynamics} we provide the exact thermodynamics of the $XY$ and Heisenberg chains. \ref{Numerical Methods} contains an account of our numerical methods for the construction of thermal states and time evolution. In \ref{Infinite Temperature} we show evidence for spin diffusion at infinite temperature, and in \ref{Energy Correlations} we show the diffusion of energy at all temperatures. 
%In \ref{Decay of Staggered Correlations} we examine the staggered correlations -- i.e. the correlations of the AFM order parameter. 
Finally, in \ref{Equilibration Dynamics} we provide more information about our equilibration simulations, and provide the extracted anomalous exponents.

%%%%%%%%%%
\section{Exact Thermodynamics \label{Exact Thermodynamics}}
%%%%%%%%%%

For reference, we provide here the exact thermodynamics of the classical Heisenberg \cite{fisher1964magnetism} and $XY$ chains. The derivation is given with open boundary conditions, since this is simpler - but the boundary effects vanish in the thermodynamic limit. 

Consider first the $XY$ chain. The partition function is

\eqa{ 
Z_{\XY} &= \int \left(\prod_{i = 1}^N \frac{d\phi_i}{2\pi}\right) \exp(\beta J \sum_{i = 1}^{N - 1} \cos(\phi_i - \phi_{i+1})) \nn \\
&= \prod_{i = 1}^{N - 1} \int \frac{d\varphi_i}{2\pi} \exp(\beta J \cos(\varphi_i)) \times \int \frac{d\varphi_N}{2\pi} \nn \\
&= I_0(\beta J)^{N-1},
}
where $I_n$ denotes a modified Bessel function of the first kind, and in the second line we have transformed the coordinates to $\varphi_i = \phi_i - \phi_{i+1}$. Similarly, the partition function of the Heisenberg chain is
\eqa{
Z &= \int \left( \prod_{i = 1}^N \frac{d\bS_i}{4\pi} \right) \exp(\beta J \sum_{i=1}^{N-1} \bS_i \cdot \bS_{i+1}) \nn \\
&= \int \frac{d\hat{\Omega}_1}{4\pi} \times \prod_{i = 2}^{N} \int \frac{d\hat{\Omega}_i}{4\pi} \exp(\beta J \cos(\theta_i)) \nn \\
&= \left( \frac{\sinh(\beta J)}{\beta J} \right)^{N-1},
}
where, in the second line, we rotate the coordinate system of $\bS_{i+1}$ such that the z-axis aligns with $\bS_i$.

From these expressions, all of the thermodynamic quantities may be calculated. In particular, taking the thermodynamic limit and setting $J = 1$, the internal energy density and specific heat are:
\eqa{
&\ep_{\XY}(T) = -I_1(1/T)/I_0(1/T), \nn \\
&C_{\XY}(T) = \frac{1}{T^2} \left(\frac{I_2(1/T) + I_0(1/T)}{2 I_0(1/T)} - \frac{I_1(1/T)^2}{I_0(1/T)^2}\right)
\label{XY_thermal}
}
for the $XY$ chain, and:
\eqa{
&\ep(T) = T - \coth(1/T), \nn \\
&C(T) = 1 - \frac{\csch(1/T)^2}{T^2}
\label{H_thermal}
}
for the Heisenberg chain. The equal-time two-point correlation function is given by
\eqn{
\avg{\bS_i \cdot \bS_j} = (-\ep)^{|i - j|} =: e^{-|i-j|/\xi},
}
and so the correlation length is
\eqn{ 
\xi(\ep) = -1/\log(-\ep),
}
which, as a function of internal energy, is the same for both the $XY$ and Heisenberg chains. In the antiferromagnet, the above is replaced with the staggered correlator.

%%%%%%%%%%
\section{Numerical Methods \label{Numerical Methods}}
%%%%%%%%%%

%%%%%
\subsection*{Construction of Thermal States}
%%%%%

Our initial thermal states are constructed using a heatbath Monte Carlo method \cite{loison2004canonical}, where we use the fact that we can precisely invert the thermal probability distribution for a single spin in a magnetic field.

The general method is known as inverse transform sampling. Let $X \in [a, b] \subseteq \mathbb{R}$ a random variable on some real interval, with probability density function $p(X)$. The cumulative distribution function (CDF) is
\eqn{
\mathcal{F}_X(x) = \int_a^x dX \;p(X),
}
i.e., the probability that a randomly sampled $X$ is less than or equal to $x$. Then the random variable $Y = \mathcal{F}^{-1}_X(u)$, where $u$ is uniformly random over $[0, 1]$, has the same probability distribution as the original variable $X$ since, by construction,
\eqn{
\mathcal{F}_Y(x) = \int_0^{u = \mathcal{F}_X(x)} du' = \mathcal{F}_X(x).
}

The specific problem is to invert the CDF. For a Heisenberg spin, this can be done analytically. Letting $\bh = h\hat{\bz}$, the CDF for $S^z$ is
\eqn{
\mathcal{F}_z(S^z) = \int^{S^z}_{-1} dz \,\frac{e^{\beta h z}}{\mathcal{Z}} = \frac{\int^{S^z}_{-1} dz\; e^{\beta h z}}{\int^1_{-1} dz \;e^{\beta h z}} = \frac{e^{\beta h S^z} - e^{-\beta h}}{e^{\beta h} - e^{-\beta h}}.
}
Then, using $\mathcal{F}_z(\mathcal{F}^{-1}_z(u)) = u$, we find that we can sample $S^z$ as
\eqn{
S^z = \mathcal{F}^{-1}_z(u) = 1 + \frac{\log(1 - u + u e^{-2 \beta h})}{\beta h}.
}
The $S^x$ and $S^y$ components have random direction in the plane, and magnitude set by the condition $|\bS| = 1$. For a magnetic field of arbitrary direction, one need only appropriately rotate the sampled spin. 

For $XY$ spins, the inversion cannot be performed analytically. In this case, we let $\bh = h\hat{\bx}$, and sample the angle $\phi$, where $(S^x, S^y) = (\cos(\phi), \sin(\phi))$, by numerically solving the integral equation
\eqn{
u = \int_{-\pi}^{\phi} d\phi' \;\frac{e^{\beta h \cos(\phi')}}{I_0(\beta h)}, 
}
for randomly generated $u$. Again, the generalisation to arbitrary magnetic field direction is via a rotation. 

To sample a state from the canonical ensemble, we randomly generate the first spin $\bS_1$. We then sweep through the chain, generating $\bS_{i+1}$ from the thermal distribution of the effective field $\bh = J\bS_i$. For open boundary conditions, cf. the coordinate transformations used to calculate the partition functions, this exactly samples the canonical ensemble. For periodic boundary conditions we perform an additional 1000 sweeps through the chain. 

%Further information on Monte Carlo heatbath algorithms may be found in \cite{loison2004canonical}.
% Including the above sentence wrecks the formatting, so I just put the reference earlier.

\iffalse
This can be done analytically for a Heisenberg spin, where the component parallel to the field of magnitude $h$ is
\eqn{
S^{||} = 1 + \frac{\log(1 - u + u e^{-2 \beta h})}{\beta h},
}
and the perpendicular component has uniformly random direction in the plane, and magnitude set by $|\bS| = 1$.
\fi

%%%%%
\subsection*{Time Evolution}
%%%%%

For the equilibration simulations, we integrate the equations of motion with the standard fourth-order Runge-Kutta (RK4) method, using a fixed timestep of $\Delta t = 0.002J^{-1}$. This method conserves the magnetisation to machine precision, and for a system size $L = 16384$ and a final time $t_f = 4096J^{-1}$ the error in the energy density is limited to $\sim 10^{-10}$. 

For the equilibrium simulations, we use a system size of $L = 8192$, but a much longer final time $t_f = 1.1 \times 10^5 J^{-1}$. To achieve such times, we use the discrete-time odd-even (DTOE) method, with a larger timestep of $\Delta t = 0.05J^{-1}$. The method consists of updating the odd spins for a timestep $\Delta t$ by exactly solving the equations of motion with the even spins held fixed, and then vice versa. This method conserves the energy to machine precision, but the error in the magnetisation is suppressed only as $O(\Delta t^2)$. However, the error does not grow with time -- for the timestep chosen the magnetisation error is $\sim 10^{-5}$.

%%%%%%%%%%
\section{Spin Diffusion at Infinite Temperature \label{Infinite Temperature}}
%%%%%%%%%%

\begin{figure}[h]
    \centering
    \includegraphics[width=\columnwidth]{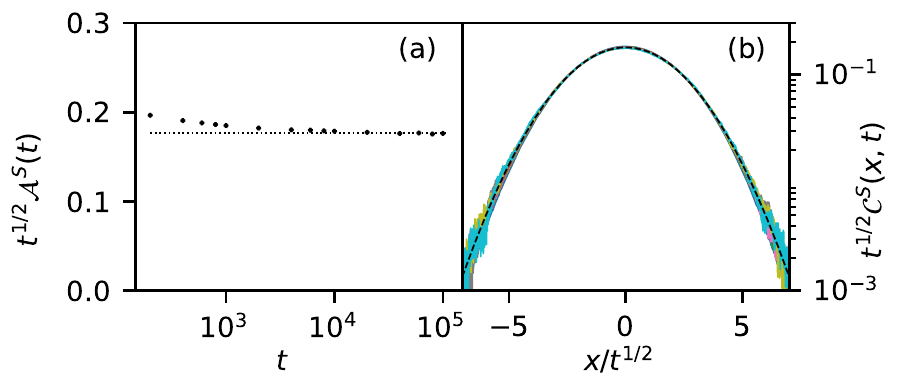}
    \caption{\small Spin diffusion at infinite temperature. Panel (a) shows the absence of the logarithmic anomaly at long times; panel (b) shows the scaling collapse. 
    \label{fig:E=0}}
\end{figure}

There has been some recent controversy over the nature of the hydrodynamics at $T = \infty$ in the Heisenberg chain, with \cite{de2020universality} claiming logarithmically enhanced diffusion and \cite{glorioso2020hydrodynamics} arguing for ordinary spin diffusion. Here we provide our own contribution to this debate: we see no evidence for logarithmically enhanced diffusion at long times, which would predict $t^{1/2} \mathcal{A}(t) \rightarrow 0$ as $t \rightarrow \infty$, see Fig.~\ref{fig:E=0}(a). 

In Fig.~\ref{fig:E=0}(b) we show the diffusive scaling collapse of the spin correlations from $t = 2000$ to $t = 10^5$. Note that the ferromagnet and antiferromagnet are indistinguishable at infinite temperature.

%%%%%%%%%%
\section{Energy Correlations \label{Energy Correlations}}
%%%%%%%%%%

\begin{figure}
    \centering
    \includegraphics[width=\columnwidth]{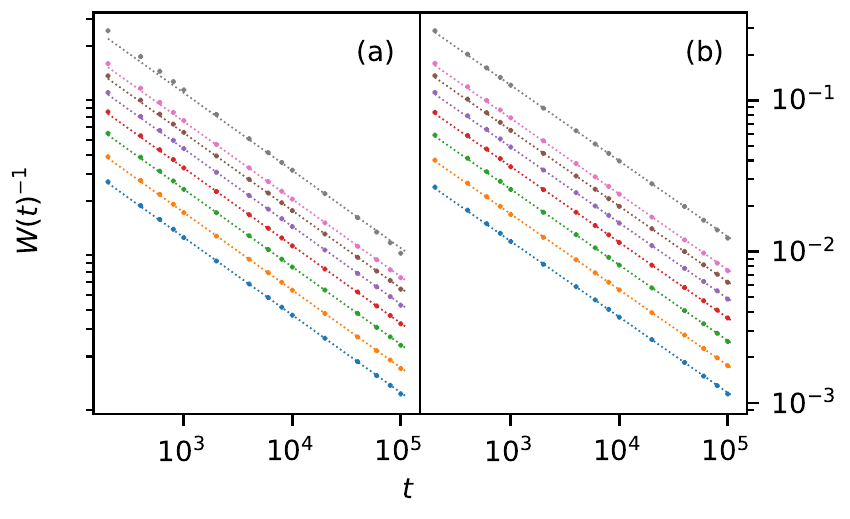}
    \caption{\small Energy diffusion in (a) the FM and (b) the AFM. The inverse-widths of $\mC^E(x,t)$ with the fit to the diffusive power-law are plotted, in ascending order, for $\ep = 0$ to $\ep = -0.7$, in steps of $-0.1$. The data are shifted vertically for clarity.
    \label{fig:E_diffusion}}
\end{figure}

We have reported in the main text that the energy correlations are found to be diffusive for both the FM and the AFM. We show the evidence for this in Fig.~\ref{fig:E_diffusion}, where we plot the Gaussian width of the energy correlations $\mC^E(x, t)$ as a function of time. We find that, except at the lowest temperatures, they are well-fit by the diffusive power-law. 

\begin{figure}
    \centering
    \includegraphics[width=\columnwidth]{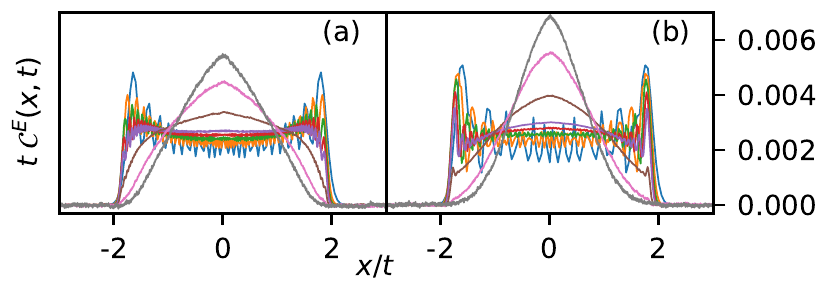}
    \caption{\small Short time ballistic propagation of the energy correlations at low temperature, $\ep = -0.9$ ($T = 0.1$), for (a) the FM and (b) the AFM. The fixed times are: $t = 20$ (blue), $t = 40$ (orange), $t = 60$ (green), $t = 80$ (red), $t = 100$ (purple), $t = 200$ (brown), $t = 400$ (pink), and $t = 600$ (grey). 
    \label{fig:E=-0.9}}
\end{figure}

At low temperatures, ballistically propagating spin-wave modes persist to intermediate times. This makes the observation of diffusion in our simulations rather difficult -- even at longer times where the ballistic modes have decayed -- because, over their lifetime, the ballistic modes increase the width of the correlations much faster than diffusion. By the time this effect is negligible, the width is comparable to the system size, and finite size effects take over. We show the ballistic collapse in Fig.~\ref{fig:E=-0.9}.% the ballistic collapse of the front.

\section{Equilibration Dynamics \label{Equilibration Dynamics}}
%%%%%%%%%

We provide here some further details of the equilibration simulations - in particular, how we determine the thermal values, and the temperature dependence of the (finite-time) anomalous exponents. 

\begin{figure}[t]
    \centering
    \includegraphics[width=7.5cm]{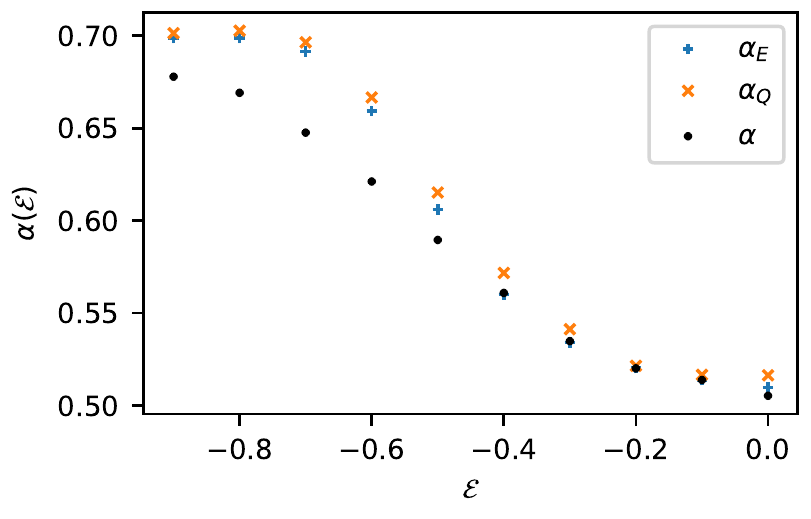}
    \caption{\small Anomalous equilibration exponents $\alpha_E$ and $\alpha_Q$ for the observables $E^{\mu}$ and $Q^{\mu}$ in the FM. The measured equilibrium exponent is also shown for comparison. The discrepancy at low-temperature is probably due to timescales -- the equilibrium exponent is extracted over the range $t = $ 10,000 to $t = $ 100,000, whereas the equilibration exponents are extracted up to $t = 4096$.}
    \label{fig:alphas_eq}
\end{figure}

We begin from a thermal state of the $XY$ chain, with every spin confined to the plane $S^z = 0$, and evolve towards a quasi-thermal state of the Heisenberg chain. Recall from the main text that we measure the degree of anisotropy with the observables
\eqn{
E^{\mu}(t) = -J\avg{S_i^{\mu}(t) S_{i + 1}^{\mu}(t)}
\label{E_mu}
}
and
\eqn{
Q^{\mu}(t) = \avg{S_i^{\mu}(t)^2}.
\label{Q_mu}
}
\iffalse
We do not consider any cross-time correlations, as we did in equilibrium, because the out-of-equilibrium correlators do not have time-translation symmetry.
\fi
For a state with energy density $\ep$, these observables are constrained by $\sum_{\mu} Q^{\mu} = 1$ and $\sum_{\mu} E^{\mu} = \ep$. Their Heisenberg equilibrium values are thus determined by isotropy, to wit, $Q^{\mu}_{\eq} = 1/3$ and $E^{\mu}_{\eq} = \ep/3$.

There is a caveat: at finite size there is a small, but conserved, total magnetisation, which prevents $Q^{\mu}$ and $E^{\mu}$ from attaining their precise equilibrium values. However, this correction may be calculated exactly. Given, at system size $L$, the $q = 0$ component of the static structure factor,
\eqn{
\Delta(L) = \frac{1}{L} \sum_{j = -L/2}^{L/2 - 1} (-\ep)^{|j|},
}
%\tb{Same problem with the indices as above}
the asymptotic values are:
\eqa{
&Q^z \rightarrow 1/3 - \Delta/3, \;\;\; \Qp \rightarrow 1/3 + \Delta/6, \nonumber \\
&E^z \rightarrow \ep/3 + J\Delta/3, \;\;\; \Ep \rightarrow \ep/3 - J\Delta/6,
}
where we have defined the averages of the in-plane observables, $\Qp = (Q^x + Q^y)/2$ and  $\Ep = (E^x + E^y)/2$.
%(\tb{what does $r$ stand for? One option would be $\perp$ and $\parallel$}) 
We take the average of the two in-plane components to account for the finite-size magnetisation spontaneously breaking the rotational symmetry of the initial $XY$ state -- which means we can read off the initial values exactly from the sum rules.

To examine the equilibration of energy fluctuations, we consider the heat capacity. Recall that the heat capacity can be estimated from a thermal ensemble as
\eqn{
C = \frac{\pd \avg{E}}{\pd T} = \frac{\avg{E^2} - \avg{E}^2}{T^2},
\label{ensemble_heat_capacity}
}
where $E$ is the energy of the state and the angle brackets denote the \textit{ensemble} average. Since the energy of each state is conserved by the Hamiltonian dynamics, this is time-independent. 

However, we define the heat capacity of a \textit{single state} as
\eqn{
C = \frac{\var{E_i}}{T^2},
}
where the variance is taken over the spatial distribution of the energy. In equilibrium, this is equal to the ensemble-based definition (\ref{ensemble_heat_capacity}), but it is not conserved by the dynamics. The initial value, of course, is the heat capacity (\ref{XY_thermal}) of the $XY$ chain, except that, since the correspondence between internal energy and temperature is different in the two chains, we must multiply the above by $(T_{\XY}/T_H)^2$, with $T_{\XY}$ and $T_H$ the temperatures that correspond to $\ep$. The equilibrium value $C_{\eq}$ is then given by the heat capacity of the Heisenberg chain (\ref{H_thermal}).

As mentioned in the main text, the equilibration simulations probe a different aspect of the underlying phenomenology: $Q^{\mu}$ and $E^{\mu}$ equilibrate diffusively in the AFM, but anomalously in the FM. The heat capacity always equilibrates diffusively.

The anomalous exponents obtained from the equilibration in the FM are shown in Fig.~\ref{fig:alphas_eq}.

\iffalse
\amcr{We may want to mention something along the lines of: 
As a matter of perspective, the observation of anomalous thermalisation is, perhaps, more robust than the anomalous scaling in equilibrium -- in the sense that, in examining the scaling, one is deliberately looking for the long-time limit; in a thermalisation experiment, one is waiting for the observables to attain their thermal values, which may, for all experimental purposes, happen before the scaling limit is approached. Indeed, this is what we observe: by the time $\delta\mO \sim 10^{-5}$, we are still in the anomalous regime.}

\amcr{In fact, we may want to make this remark in the main text...}
\fi

%\bibliography{refs}
%\input{main.bbl}

\end{document}